\documentclass[11pt,english]{article}
\usepackage[T1]{fontenc}
\usepackage{a4wide}
\usepackage{babel}

\makeatletter

\providecommand{\LyX}{L\kern-.1667em\lower.25em\hbox{Y}\kern-.125emX\@}

\makeatother

\begin{document}

\begin{flushright}CBPF-NF-038/02\\November 2002 

\end{flushright} 

\vspace{1cm}

\begin{center}

\textbf{\LARGE \( Z_{k} \) string fluxes and monopole confinement in non-Abelian
theories}{\LARGE \par}

\vspace{1cm}

{\large {\bf Marco A. C. Kneipp}}\footnote{e-mail: {\tt kneipp@cbpf.br}} 

\vspace{0.3cm}

\end{center}

\begin{center}{\em Universidade do Estado do Rio de Janeiro(UERJ),\\

Depto. de Física Teórica,\\

Rua São Francisco Xavier, 524\\

20550-013, Rio de Janeiro, Brazil.\\
\medskip{}

Centro Brasileiro de Pesquisas F\' \i sicas (CBPF), \\ 

Coordenação de Teoria de Campos e Partículas (CCP), \\ Rua Dr. Xavier Sigaud,
150 \\

22290-180, Rio de Janeiro, Brazil. \\ }

\end{center}

\begin{abstract}
Recently we considered \( N=2 \) super Yang-Mills with a \( N=2 \) mass breaking
term and showed the existence of BPS \( Z_{k} \)-string solutions for arbitrary
simple gauge groups which are spontaneously broken to non-Abelian residual gauge
groups. We also calculated their string tensions exactly. In doing so, we have
considered in particular the hypermultiplet in the same representation as the
one of a diquark condensate. In the present work we analyze some of the different
phases of the theory and find that the magnetic fluxes of the monopoles are
multiple of the fundamental \( Z_{k} \)-string flux, allowing for monopole
confinement in one of the phase transitions of the theory. We also calculate
the threshold length for a string breaking. Some of these confining theories
can be obtained by adding a \( N=0 \) deformation term to \( N=2 \) or \( N=4 \)
superconformal theories.

\vfill PACS numbers: 11.27.+d, 11.15.-q, 02.20.-a, 12.38.AW.
\end{abstract}

\section{Introduction}

It is long believed that the quark confinement would be dual to a non-Abelian
generalization of Meissner effect, as proposed by 't Hooft and Mandelstam many
years ago\cite{tHooftMandelstam}. Following their ideas an important progress
has been made by Seiberg and Witten \cite{SeibergWitten1} which, starting from
an \( N=2 \) \( SU(2) \) supersymmetric theory, obtained an effective \( N=2 \)
\( U(1) \) super QED with an \( N=2 \) mass breaking term associated to the
point in the moduli space where a monopole becomes massless. In this effective
theory, the \( U(1) \) group is broken to a discrete subgroup. As it happens
the theory develops (solitonic) string solutions and monopoles confinement occurs,
with these monopoles being identified with electric charges. After that, many
interesting works appeared \cite{varios}, analyzing various aspects of \( N=2 \)
\( SU(N_{c}) \) SQCD with, \( N=2 \) \( U(1)^{N_{c}-1} \)theory with a \( N=2 \)
mass breaking term as effective theory. These effective theories also have string
solutions when the gauge group is broken by Higgs mechanism. These string solutions
confine monopoles and carry topological charges in the group \( \Pi _{1}\left[ U(1)^{N_{c}-1}/Z^{N_{c}-1}\right] =Z^{N_{c}-1} \).
On the other hand, recently we considered \cite{KB} \( N=2 \) super Yang-Mills
with a \( N=2 \) mass breaking term, with arbitrary simple gauge group \( G \)
which in general is broken to a non-Abelian residual gauge group. One of the
spontaneous symmetry breaking is produced by a complex scalar \( \phi  \) that
could be in the symmetric part of the tensor product of \( k \) fundamental
representations, with \( k\geq 2 \). In particular if \( k=2 \), this is the
same representation as the one of a diquark\footnote{%
By quark we mean a fermion in a fundamental representation.
} condensate and therefore this scalar can be thought as being itself the diquark
condensate (or the monopole condensate in the dual theory). Therefore, when
\( k=2 \), we can consider this theory as being an effective theory, like the
Abelian-Higgs is an effective theory for BCS theory. Besides the fact that our
``effective'' theory has a non-Abelian gauge group \( G \), another interesting
feature is that it has solitonic monopoles which are solutions of the equations
of motion, differently from the Dirac monopoles which usually appear in the
Abelian theories. When the scalar \( \phi  \) acquires an expectation value
it breaks the gauge group \( G \) into \( G_{\phi } \), such that \( \Pi _{1}(G/G_{\phi })=Z_{k} \),
allowing the existence of \( Z_{k} \)-strings solutions. We have showed the
existence of BPS \( Z_{k} \)-string solutions for these theories (for arbitrary
\( k\geq 2 \)) and calculated exactly their string tensions. In the present
work, we analyze some other features of these theories. In section 2 we show
that by varying continuously a mass parameter \( m \) we can pass from an unbroken
phase to a phase with free monopoles and then to a phase with \( Z_{k} \)-strings
and confined monopoles. In section 3 we analyze the phase which has solitonic
BPS monopole solutions which we call ``free-monopole phase''. These monopole
solutions are expected to fill irreducible representations of the dual unbroken
gauge group\cite{HolloDoFraMACK}. In this phase \( N=2 \) supersymmetry is
recovered and we show that some of these theories are conformal. In section
4 we analyze the magnetic fluxes of the BPS strings which appear in the superconducting
phase. We show that the magnetic fluxes of the magnetic monopoles are multiple
of the fundamental string flux and therefore the monopoles can get confined.
We also obtain the threshold length of a string to break in a new pair of monopole-antimonopole.
These results are obtained considering the theory in the weak coupling regime
since, in the ``dual Meissner effect picture'' for confinement, one wants
to map a theory in the weak coupling regime with monopole confinement to a theory
in the strong coupling regime with quark confinement, through a duality transformation.
The general topological aspects for monopole confinement during a phase transition
have been given in \cite{B81} (see also \cite{PV}). Our aim here is to analyze
the monopole confinement in our specific theory. We conclude with a summary
of the results.

\section{Phases of the theory}

As is quite well known, in the broken phase of the Abelian-Higgs theory in 3+1
dimensions, there exist string solutions with string tension satisfying the
inequality
\begin{equation}
\label{1.1}
T\geq \frac{q_{\phi }}{2}a^{2}|\Phi _{\textrm{st}}|\, ,
\end{equation}
where \( a \) is a breaking parameter which appears in the potential, \( \Phi _{\textrm{st}} \)
is the string's magnetic flux which satisfies the quantization condition
\begin{equation}
\label{1.2}
\Phi _{\textrm{st}}=\frac{2\pi n}{q_{\phi }}\, ,\, \, n\in Z
\end{equation}
and \( q_{\phi } \) is the electric charge of the scalar field. Considering
that \( \phi ^{\dagger } \) is a condensate of electron pairs, then\footnote{%
Considering that \( \hbar =1=c \) 
} \( q_{\phi }=2e \). Following 't Hooft and Mandelstam's \cite{tHooftMandelstam}
idea, if one considers a Dirac monopole-antimonopole system in the Abelian-Higgs
theory, the magnetic lines can not spread over space but must rather form a
string which gives rise to a confining potential between the monopoles. That
is only possible because the Dirac monopole magnetic flux is \( \Phi _{\textrm{mon}}=g=2\pi /e \),
which is twice the fundamental string's magnetic flux, allowing one to attach
to the monopole two strings with \( n=1 \). 

One of our aims in this work is to generalize some of these ideas on monopole
confinement to non-Abelian gauge theories. For simplicity let us consider a
gauge group \( G \) which is simple, connected and simply-connected, and adopt
the same conventions as in \cite{KB}. Following our previous work, we shall
consider a Yang-Mills theory with a complex scalar \( S \) in the adjoint representation
and a complex scalar \( \phi  \) in another particular representation. We consider
a scalar in the adjoint representation because in a spontaneous symmetry breaking
it can produce an unbroken gauge group with a \( U(1) \) factor, which allows
the existence of monopole solutions. Additionally, another motivation for having
a scalar in the adjoint representation is because with it, we can form an \( N=2 \)
vector supermultiplet and, like in the Abelian-Higgs theory, the BPS string
solutions appear naturally in a theory with \( N=2 \) supersymmetry and a \( N=2 \)
mass breaking term. Moreover, in a theory with the field content of \( N=2 \),
the monopole spin is consistent with the quark-monopole duality\cite{Osborn}
which is another important ingredient in 't Hooft and Mandelstam's ideas. In
order to have monopole confinement we need also to have string solutions. A
necessary condition for the existence of a (topological) string is to have a
non-trivial first homotopy group of the vacuum manifold. One way to produce
a spontaneous symmetry breaking satisfying this condition is to introduce a
complex scalar \( \phi  \) in a representation which contains the weight state
\( \left| k\lambda _{\phi }\right\rangle  \) \cite{OT}, where \( k \) is
an integer greater or equal to two, and \( \lambda _{\phi } \) a fundamental
weight. For an arbitrary gauge group \( G \) there are at least three possible
representations which have this weight state: one is to consider \( \phi  \)
in the representation with \( k\lambda _{\phi } \) as highest weight, which
we shall denote \( R_{k\lambda _{\phi }} \). We can also consider \( \phi  \)
to be in the direct product of \( k \) fundamental representations with fundamental
weight \( \lambda _{\phi } \), which we shall denote \( R_{k\lambda _{\phi }}^{\otimes } \).
Finally a third possibility would be to consider \( \phi  \) in the symmetric
part of \( R_{k\lambda _{\phi }}^{\otimes } \), called \( R_{k\lambda _{\phi }}^{\textrm{sym}} \),
which always contains \( R_{k\lambda _{\phi }} \). This last possibility has
an extra physical motivation that if \( k=2 \), it corresponds to the representation
of a condensate of two massless fermions (in the microscopic theory) in the
fundamental representation with fundamental weight \( \lambda _{\phi } \) which
we shall loosely call quarks. Therefore, for \( k=2 \), we could interpret
\( \phi  \) as being this diquark condensate. In this case, when \( \phi  \)
takes a non-trivial expectation value, it also gives rise to a mass term for
these quarks. In order to have \( N=2 \) supersymmetry we should need another
complex scalar to be in the same hypermultiplet as \( \phi  \). For simplicity's
sake, however, we shall ignore it setting it to zero. Note for the gauge group
\( SU(n) \), the scalar \( S \) in the adjoint representation could also be
interpreted as a bound state of quark-antiquark, for the quark in the \( n \)
dimensional representation.

Let us consider the Lagrangian used in \cite{KB}, 
\begin{equation}
\label{Lagrangean}
L=-\frac{1}{4}G^{\mu \nu }_{a}G_{a\mu \nu }+\frac{1}{2}\left( D^{\mu }S\right) _{a}^{*}\left( D_{\mu }S\right) _{a}+\frac{1}{2}\left( D^{\mu }\phi \right) ^{\dagger }D_{\mu }\phi -V(S,\phi )
\end{equation}
with
\begin{equation}
\label{2.1}
V(S,\phi )=\frac{1}{2}\left( Y_{a}^{2}+F^{\dagger }F\right) \geq 0
\end{equation}
where 
\begin{eqnarray}
Y_{a} & = & \frac{e}{2}\left\{ \left( \phi ^{\dagger }T_{a}\phi \right) +S^{*}_{b}if_{bca}S_{c}-m\left( \frac{S_{a}+S_{a}^{*}}{2}\right) \right\} \, ,\label{2.2a} \\
F & \equiv  & e\left( S^{\dagger }-\frac{\mu }{e}\right) \phi \, .\label{2.2b} 
\end{eqnarray}
\( T_{a} \) are the orthogonal Lie algebra generators which satisfy
\begin{equation}
\label{2.3}
\textrm{Tr}\, \left( T_{a}T_{b}\right) =x_{\phi }\psi ^{2}\delta _{ab}
\end{equation}
where \( x_{\phi } \) is the Dynkin index of \( \phi  \)'s representation
and \( \psi ^{2} \) is the length square of the highest root which we shall
take to be 2. This Lagrangian is the bosonic part of \( N=2 \) super Yang-Mills
with one flavor (with one of the aforementioned scalars of the hypermultiplet
put equal to zero) and a \( N=2 \) breaking mass term \footnote{%
One can check that easily by comparing with the Lagrangian of \( N=2 \) super
Yang-Mills written in the appendix of \cite{KB}.
}. The parameter \( \mu  \) gives a bare mass to \( \phi  \) and \( m \) gives
a bare mass to the \textit{real part} of \( S \) which therefore a \( N=0 \)
deformation of a \( N=2 \) SQCD. This breaking is different from the one considered
by Seiberg-Witten \cite{SeibergWitten1}, which breaks \( N=2 \) to \( N=1 \).
We shall consider that this theory is in the weak coupling regime.

The vacua are solutions to equation \( V(S,\phi )=0 \), which is equivalent
to the conditions
\begin{equation}
\label{2.4}
Y_{a}=0=F\, .
\end{equation}
In order to the topological string solutions to exist, we look for vacuum solutions
of the form 
\begin{eqnarray}
\phi ^{\textrm{vac}} & = & a|k\lambda _{\phi }>\, ,\label{2.5a} \\
S^{\textrm{vac}} & = & b\lambda _{\phi }\cdot H\, ,\label{2.5b} 
\end{eqnarray}
where \( a \) is a complex constant, \( b \) is a real constant and \( |k\lambda _{\phi }> \)
is a weight state with \( \lambda _{\phi } \) being an arbitrary fundamental
weight and \( k \) being an integer greater or equal to two. If \( a\neq 0 \),
this configuration breaks \( G\rightarrow G_{\phi } \) in such a way that\cite{OT}
\( \Pi _{1}(G/G_{\phi })=Z_{k} \), which is a necessary condition for the existence
of \( Z_{k}- \)strings. Let us consider that \( \mu >0 \). Following \cite{KB},
from the vacuum conditions (\ref{2.4}) one can conclude that 
\begin{eqnarray*}
|a|^{2} & = & \frac{mb}{k}\, ,\\
\left( kb\lambda _{\phi }^{2}-\frac{\mu }{e}\right) a & = & 0\, .
\end{eqnarray*}
 There are three possibilities:

\begin{description}
\item [(i)]\( m<0\, \, \Rightarrow  \) If \( a\neq 0 \), then \( b=\mu /\lambda _{\phi }^{2}ke>0 \)
which would imply \( |a|^{2}<0 \). Therefore we must have \( a=0=b \) and
the gauge group \( G \) remains unbroken.
\item [(ii)]\( m=0\, \, \Rightarrow \, \, a=0 \) and \( b \) can be an arbitrary
constant. In this case, considering \( b\neq 0 \), \( S^{\textrm{vac}} \)breaks\cite{OT}
\begin{equation}
\label{2.6}
G\rightarrow G_{S}\equiv \left( K\times U(1)\right) /Z_{l}\, ,
\end{equation}
 where \( K \) is the subgroup of \( G \) associated to the algebra whose
Dynkin diagram is given by removing the dot corresponding to \( \lambda _{\phi } \)
from that of \( G. \) The \( U(1) \) factor is generated by \( \lambda _{\phi }\cdot H \)
and \( Z_{l} \) is a discrete subgroup of \( U(1) \) and \( K \). The order
\( l \) of \( Z_{l} \) is equal to \( p_{\phi }|Z(K)|/|Z(G)| \) where \( |Z(G)| \)
and \( |Z(K)| \) are respectively the orders of the centers of the groups \( G \)
and \( K \) and \( p_{\phi } \) is the smallest integer such that \( p_{\phi }2\lambda _{\phi }/\alpha ^{2}_{\phi } \)
is in the coroot lattice\cite{GodOlive(81)}.
\item [(iii)]\( m>0\, \Rightarrow  \) Besides the solution \( a=0=b \), we can have
\begin{eqnarray}
|a|^{2} & = & \frac{m\mu }{k^{2}e\lambda _{\phi }^{2}}\, ,\label{2.7a} \\
b & = & \frac{\mu }{ke\lambda _{\phi }^{2}}\, ,\label{2.7b} 
\end{eqnarray}
and \( G \) is further broken to\cite{OT}
\begin{equation}
\label{2.8}
G\rightarrow G_{\phi }\equiv \left( K\times Z_{kl}\right) /Z_{l}\supset G_{S}\, .
\end{equation}
In particular, for \( k=2 \), we can have for example the symmetry breaking
patterns, 
\begin{eqnarray*}
\textrm{Spin}(10) & \rightarrow  & \left( SU(5)\times Z_{10}\right) /Z_{5}\, ,\\
SU(3) & \rightarrow  & \left( SU(2)\times Z_{4}\right) /Z_{2}\, .
\end{eqnarray*}

\end{description}
Therefore by continuously changing the value of the parameter \( m \) we can
produce a symmetry breaking pattern \( G\, \rightarrow \, G_{S}\, \rightarrow \, G_{\phi } \).
It is interesting to note that, unlike the Abelian-Higgs theory, in our theory
the bare mass \( \mu  \) of \( \phi  \) is not required to satisfy \( \mu ^{2}<0 \)
in order to have spontaneous symmetry breaking. Therefore in the dual formulation,
where one could interpret \( \phi  \) as being the monopole condensate, we
don't need to have a monopole mass satisfying the problematic condition \( M_{\textrm{mon}}^{2}<0 \)
mentioned by 't Hooft\cite{tHooftreview}. 

Let us analyze in more detail the last two phases.

\section{The \protect\( m=0\protect \) or free-monopole phase}

When \( m=0 \), \( N=2 \) supersymmetry is restored. In this phase \( a=0 \)
and \( b \) is an arbitrary constant, which we shall consider it to be given
by (\ref{2.7b}), in order to have the same value as the case when \( m<0 \).
The vacuum configuration \( S^{\textrm{vac}} \) defines the \( U(1) \) direction
in \( G_{S} \), (\ref{2.6}), and one can define the corresponding \( U(1) \)
charge as \cite{GOrev}
\begin{equation}
\label{3.1}
Q\equiv e\frac{S^{\textrm{vac}}}{|S^{\textrm{vac}}|}=e\frac{\lambda _{\phi }\cdot H}{|\lambda _{\phi }|}\, .
\end{equation}
Since in this phase \( \Pi _{2}(G/G_{S})=Z \), it can exist \( Z \)-magnetic
monopoles. These solutions can be written in the following form\cite{B78}:
for each root \( \alpha  \), such that \( 2\alpha ^{\textrm{v}}\cdot \lambda _{\phi }\neq 0 \)
(where \( \alpha ^{\textrm{v}}\equiv \alpha /\alpha ^{2} \)), we can define
the generators
\begin{equation}
\label{3.2}
T^{\alpha }_{1}=\frac{E_{\alpha }+E_{-\alpha }}{2}\, \, ,\, \, \, T_{2}^{\alpha }=\frac{E_{\alpha }-E_{-\alpha }}{2i}\, \, ,\, \, \, T_{3}^{\alpha }=\frac{\alpha \cdot H}{\alpha ^{2}}
\end{equation}
which satisfy the SU(2) algebra
\[
\left[ T_{i}^{\alpha },T_{j}^{\alpha }\right] =i\epsilon _{ijk}T_{k}^{\alpha }\, .\]
Using spherical coordinates we define the group elements
\begin{equation}
\label{3.3}
g^{\alpha }_{p}(\theta ,\phi )\equiv \exp \left( ip\varphi T_{3}^{\alpha }\right) \exp \left( i\theta T^{\alpha }_{2}\right) \exp \left( -ip\varphi T^{\alpha }_{3}\right) \, ,\, \, \, \, \, p\in Z\, .
\end{equation}
Let \( S=M+iN \), where \( M \) and \( N \) are real scalar fields. The asymptotic
form for the scalars of the \( Z- \)monopole are obtained by performing a gauge
transformation on the vacuum solution (\ref{2.5a}), (\ref{2.5b}) by the above
group elements. This results, at \( r\rightarrow \infty  \) 
\begin{eqnarray}
M(\theta ,\phi ) & = & g^{\alpha }_{p}v\cdot H\left( g^{\alpha }_{p}\right) ^{-1}\, ,\label{3.4a} \\
N(\theta ,\phi ) & = & 0\, \, \, ,\, \, \, \, \, \phi (\theta ,\phi )=0\, .\label{3.4c} 
\end{eqnarray}
 where \( v\equiv b\lambda _{\phi } \). The U(1) magnetic charge of these monopoles
are \cite{B78}
\begin{equation}
\label{3.6}
g\equiv \frac{1}{|v|}\int dS_{i}M^{a}B_{i}^{a}=\frac{4\pi }{e}\frac{pv\cdot \alpha ^{\textrm{v}}}{|v|}
\end{equation}
where \( B_{i}^{a}\equiv -\epsilon _{ijk}G_{jk}^{a}/2 \) are the non-Abelian
magnetic fields. Due to the \( N=2 \) supersymmetry, these monopoles must be
BPS and satisfy the mass formula
\begin{equation}
\label{3.6a}
m_{\textrm{mon}}=|v||g|\, .
\end{equation}
Not all of these monopoles are stable. The stable or fundamental BPS monopoles
are those which \( p=1 \) and \( 2\alpha ^{\textrm{v}}\cdot \lambda _{\phi }=\pm 1 \)\cite{EWeinberg}.
From now on we shall only consider these fundamental monopoles, which are believed
to fill representations of the gauge subgroup \( K \)\cite{HolloDoFraMACK}.

It is interesting to note that for the particular case where the gauge group
is \( G=SU(2) \) and \( \phi  \) is in the symmetric part of the tensor product
of two fundamental representations, which correspond to the adjoint representation,
the supersymmetry of the theory is enhanced to \( N=4 \), and the theory has
vanishing \( \beta  \) function. There are other examples of vanishing \( \beta  \)
functions when \( m=0 \). In order to see that we must recall that the \( \beta  \)
function of \( N=2 \) super Yang-Mills with a hypermultiplet is given by
\[
\beta (e)=\frac{-e^{3}}{\left( 4\pi \right) ^{2}}\left[ h^{\textrm{v}}-x_{\phi }\right] \]
where \( h^{\textrm{v}} \) is the dual Coxeter number of \( G \) and \( x_{\phi } \)
is the Dynkin index of \( \phi  \)'s representation(\ref{2.3}). If \( \phi  \)
belongs to \( R^{\textrm{sym}}_{2\lambda _{\phi }} \),
\[
x_{\phi }=x_{\lambda _{\phi }}\left( d_{\lambda _{\phi }}+2\right) \, .\]
where \( x_{\lambda _{\phi }} \) and \( d_{\lambda _{\phi }} \) are, respectively,
the Dynkin index and the dimension of the representation associated to the fundamental
weight \( \lambda _{\phi } \). On the other hand if \( \phi  \) belongs to
the direct product of two fundamental representations, \( R^{\otimes }_{2\lambda _{\phi }} \),
\[
x_{\phi }=2d_{\lambda _{\phi }}x_{\lambda _{\phi }}\, ,\]

Therefore for \( SU(n) \) (which has \( h^{\textrm{v}}=n \)), if \( \phi  \)
is in the tensor product of the fundamental representation of dimension \( d_{\lambda _{n-1}}=n \)
with itself (which has Dynkin index \( x_{\lambda _{n-1}}=1/2 \)), then \( x_{\phi }=n \)
and the \( \beta  \) function vanishes. Therefore in this phase the theory
is \( N=2 \) superconformal (if we take \( \mu =0 \)) and \( SU(n) \) is
broken to \( U(n-1)\sim [SU(n-1)\otimes U(1)]/Z_{n-1} \).

\section{The \protect\( m>0\protect \) or superconducting phase}

In the ``\( m>0 \)'' phase, the \( U(1) \) factor of \( G_{S} \) (eq.(\ref{2.6}))
is broken and, like the Abelian-Higgs theory, the magnetic flux lines associated
to this \( U(1) \) factor cannot spread over space. Since \( G \) is broken
in such a way that \( \Pi _{1}(G/G_{\phi })=Z_{k} \), these flux lines may
form topological \( Z_{k}- \)strings. We indeed showed in \cite{KB} the existence
of BPS \( Z_{k} \)-strings in the limit \( m\rightarrow 0_{+} \) and \( \mu \rightarrow \infty  \),
with \( m\mu =\textrm{const} \). We shall show now that, as in the Abelian-Higgs
theory, the \( U(1) \) magnetic flux \( \Phi _{\textrm{mon}} \) of the above
monopoles is a multiple of the fundamental \( Z_{k} \)-string magnetic flux,
and therefore these \( U(1) \) flux lines coming out of the monopole can be
squeezed into \( Z_{k} \)-strings, which gives rise to a confining potential.

\subsection{\protect\( Z_{k}\protect \)-string magnetic flux}

From (\ref{2.5a}) and (\ref{3.1}) it follows, 
\[
Q\phi ^{\textrm{vac}}=ek|\lambda _{\phi }|\, \phi ^{\textrm{vac}}\, ,\]
and therefore the \( U(1) \) electric charge of \( \phi ^{\textrm{vac}} \)
is 
\begin{equation}
\label{4.1}
q_{\phi }=ek|\lambda _{\phi }|\, .
\end{equation}
On the other hand, the string tension satisfies the bound \cite{KB}
\begin{eqnarray}
T & \geq  & \frac{me}{2}\left| \int d^{2}x\, M^{a}B^{a}_{3}\right| =\frac{me|v|}{2}|\Phi _{\textrm{st}}|\nonumber \\
 & = & \frac{q_{\phi }}{2}|a|^{2}|\Phi _{\textrm{st}}|\label{4.2} 
\end{eqnarray}
where \( B_{i}^{a}\equiv -\epsilon _{ijk}G^{aij}/2 \) is the non-Abelian magnetic
field and 
\begin{equation}
\label{4.3}
\Phi _{\textrm{st}}\equiv \frac{1}{|v|}\int d^{2}x\, M^{a}B^{a}_{3}
\end{equation}
is the \( U(1) \) string magnetic flux, with the integral taken over the plane
perpendicular to the string. This flux definition is gauge invariant and consistent
with the flux definition for the monopole (\ref{3.6}). One notes that (\ref{4.2})
is very similar to the Abelian result (\ref{1.1}), but here \( q_{\phi } \)
and \( a \) are given by (\ref{4.1}) and (\ref{2.7a}) respectively. Let us
use the string ansatz in \cite{KB}:
\begin{eqnarray}
\phi (\varphi ,\rho ) & = & f(\rho )e^{i\varphi L_{n}}a|k\lambda _{\phi }>\, ,\nonumber \\
mS(\varphi ,\rho ) & = & h(\rho )k|a|^{2}e^{i\varphi L_{n}}\lambda _{\phi }\cdot He^{-i\varphi L_{n}}\, ,\label{4.4} \\
W_{i}(\varphi ,\rho ) & = & g(\rho )L_{n}\frac{\epsilon _{ij}x^{j}}{e\rho ^{2}}\, ,\, \, \, \, \, i,j=1,2\, \, \, \, \, \rightarrow \, \, \, \, \, B_{3}(\varphi ,\rho )=\frac{L_{n}}{e\rho }g'(\rho )\, ,\nonumber \\
W_{0}(\varphi ,\rho ) & = & W_{3}(\varphi ,\rho )=0\, ,\nonumber 
\end{eqnarray}
with the boundary conditions
\[
f(\infty )=g(\infty )=h(\infty )=1\, ,\]
\[
f(0)=g(0)=0\, ,\]
 and considering 
\[
L_{n}=\frac{n}{k}\frac{\lambda _{\phi }\cdot H}{\lambda _{\phi }^{2}}\, ,\, \, \, \, \, \, n\in Z_{k}\]
Then, using the BPS conditions obtained in \cite{KB}, which are valid in the
limit \( m\rightarrow 0 \) and \( \mu \rightarrow \infty  \), results that
the functions \( f(\rho ) \) and \( g(\rho ) \) satisfy the same differential
equations as the BPS strings in the \( N=2 \) Abelian-Higgs theory. However
that fact doesn't mean that BPS \( Z_{k} \) strings are solutions of the \( N=2 \)
Abelian-Higgs theory, since in this limit our theory continue to be non-Abelian.
Moreover, from the asymptotic configuration we obtain that the ``topological
classes'' are determined by the first homotopy group \( \Pi _{1}(G/G_{\phi }) \)
which is different from the one of the Abelian-Higgs theory. 

From the BPS condition \( D_{\pm }S=0 \) together with the boundary conditions,
results that \( h(\rho )=1 \). Therefore we obtain that for the BPS \( Z_{k} \)-strings,
\begin{equation}
\label{4.4c}
\Phi _{\textrm{st}}=\oint dl_{i}A_{i}=\frac{2\pi n}{q_{\phi }}\, \, \, \, ,\, \, \, \, \, \, \, \, \, \, n\in Z_{k\, },
\end{equation}
where \( A_{i}\equiv W_{i}^{a}M^{a}/|v|\, ,\, \, \, i=1,2 \). This flux quantization
condition is also very similar to the Abelian result (\ref{1.2}), but different
due to the value of the electric charge \( q_{\phi } \) given by (\ref{4.1}).
This result generalizes, for example, the string magnetic flux for \( SU(2) \)\cite{Schaposnik}
and for \( SO(10) \)\cite{Everett}(up to a factor of \( \sqrt{2} \) ). In
\cite{konishi}, it is also calculated the magnetic fluxes for the \( SU(n) \)
theory, but with the gauge group completely broken to its center and a different
definition of string flux which is not gauge invariant. We can also rewrite
the above result as
\[
\Phi _{\textrm{st}}q_{\phi }=2\pi n\, \, ,\, \, \, \, \, \, n\in Z_{k}\, ,\]
 which is similar to the magnetic monopole charge quantization condition.

Let us now check that the magnetic flux \( \Phi _{\textrm{mon}} \) of the monopoles
in the \( U(1) \) direction generated by \( \lambda _{\phi }\cdot H \) is
multiple of \( \Phi _{\textrm{st}} \). From (\ref{3.6}), using (\ref{4.1})
and the fact that 
\[
\alpha ^{\textrm{v}}=\sum ^{r}_{i=1}m_{i}\alpha ^{\textrm{v}}_{\textrm{i }}\, \, ,\, \, \, \, \, \, \, \alpha _{i}^{\textrm{v}}=\frac{\alpha _{i}}{\alpha ^{2}_{i}}\, \, ,\, \, \, \, \, \, m_{i}\in Z,\]
where \( \alpha _{i} \) are simple roots. It then follows that 
\[
\Phi _{\textrm{mon}}=g=\frac{2\pi km_{\phi }p}{q_{\phi }}\, .\]
Therefore, for the fundamental monopoles, which have \( p=1 \) and \( m_{\phi }=1 \),
\( \Phi _{\textrm{mon}} \) is equal to the flux \( \Phi _{\textrm{st}} \)
of the string with \( n=k \) or \( k \) strings with \( n=1 \). This can
be interpreted that for one fundamental monopole one can attach \( k \) \( Z_{k} \)-strings
with \( n=1 \). This is consistent with the fact that a set of \( k \) \( Z_{k}- \)strings
with \( n=1 \) belongs to the trivial first homotopy of the vacuum manifold
and therefore can terminate in a magnetic monopole, which also belongs to the
trivial first homotopy group. It is important to stress the fact that been in
the trivial topological sector doesn't mean that the set of \( k \) \( Z_{k}- \)strings
with \( n=1 \) has vanishing flux \( \Phi _{\textrm{st}} \).

\subsection{Monopole confinement}

In the \( m>0 \) phase, it is expected \cite{B81} that the monopoles produced
in the \( m=0 \) phase develop a flux line or string and get confined. Naively
we could see this fact in the following way: as usual, in order to obtain the
asymptotic scalar configuration of a monopole, starting from the vacuum configuration
(\ref{2.5a}) and (\ref{2.5b}) one performs the gauge transformation (\ref{3.3})
and obtains that at\footnote{%
Note that when we take \( m=0\, \Rightarrow \, a=0 \) we recover the asymptotic
scalar field configuration for the \( Z \)-monopole in the ``m=0 phase''
(\ref{3.4a}), (\ref{3.4c})
} \( r\rightarrow \infty  \),
\begin{eqnarray}
S(\theta ,\varphi ) & = & g_{p}^{\alpha }b\lambda _{\phi }\cdot H\left( g_{p}^{\alpha }\right) ^{-1}\label{4.7a} \\
\phi (\theta ,\varphi ) & = & g_{p}^{\alpha }a|k\lambda _{\phi }>\label{4.7b} 
\end{eqnarray}
 However, \( \phi (\theta ,\varphi ) \) is singular. In order to see this let's
consider for simplicity \( p=1 \), \( k=2 \), and \( \alpha  \) to be those
positive roots such that \( 2\lambda _{\phi }\cdot \alpha ^{\textrm{v}}=1 \)
. In this case, the orthonormal weight states
\[
|2\lambda _{\phi }>\, \, \, ,\, \, \, |2\lambda _{\phi }-\alpha >\, \, \, ,\, \, \, |2\lambda _{\phi }-2\alpha >\]
form a spin 1 irrep of the \( su(2) \) algebra (\ref{3.2}) and the orthonormal
states
\begin{eqnarray}
|\pm > & \equiv  & \frac{1}{2}\left( |2\lambda _{\phi }>\pm i\sqrt{2}|2\lambda _{\phi }-\alpha >-|2\lambda _{\phi }-2\alpha >\right) \, ,\label{a.1} \\
|0> & \equiv  & \frac{1}{\sqrt{2}}\left( |2\lambda _{\phi }>+|2\lambda _{\phi }-2\alpha >\right) \, ,\label{a.2} 
\end{eqnarray}
 satisfy
\begin{eqnarray*}
T^{\alpha }_{2}|\pm > & = & \pm |\pm >\, ,\\
T^{\alpha }_{2}|0> & = & 0\, .
\end{eqnarray*}
 We can then write 
\[
|2\lambda _{\phi }>=\frac{1}{2}\left( |+>+|->+\sqrt{2}|0>\right) \, .\]
Then, (\ref{4.7b}) can be written as

\[
\phi (\theta ,\varphi )=a\left\{ \cos ^{2}\frac{\theta }{2}|2\lambda _{\phi }>-\frac{\sqrt{2}}{2}\sin \theta e^{-i\varphi }|2\lambda _{\phi }-\alpha >+\sin ^{2}\frac{\theta }{2}e^{-2i\varphi }|2\lambda _{\phi }-2\alpha >\right\} \, .\]
Therefore at \( \theta =\pi  \),
\[
\phi (\pi ,\varphi )=ae^{-2i\varphi }|2\lambda _{\phi }-2\alpha >\]
 which is singular. This generalizes Nambu's result \cite{Nambu} for the \( SU(2)\times U(1) \)
case. In order to cancel the singularity we should attach a string in the \( z<0 \)
axis with a zero in the core, similar to our string ansatz (\ref{4.4}). One
could construct an ansatz for \( \phi (r,\theta ,\phi ) \) by multiplying the
above asymptotic configuration by a function \( F(r,\theta ) \) such that \( F(r,\pi )=0 \).

Since our theory have solitonic monopoles with masses given by (\ref{3.6a}),
we can obtain a bound for the threshold length for a string to break producing
a new monopole-antimonopole pair in the following way: from (\ref{4.2}) and
(\ref{4.4c}), it follows that the string tension for a string with \( n=k \)
or \( k \) strings with \( n=1 \) satisfy\footnote{%
Like in the Abelian-Higgs theory, it is expected that when \( V(\phi \, S)\geq Y_{a}^{2}/2 \),
as we are considering, the tension of a string with \( n=k \) should be greater
or equal to the tension of \( k \) strings with \( n=1 \).
} the bound 
\[
T\geq k\frac{me|v|}{2}\frac{2\pi }{q_{\phi }}\, .\]
 Using the monopole masses (\ref{3.6a}), the threshold length \( d^{\textrm{th}} \)
for a string breaking can be derived form the relation
\[
2|v|\frac{2\pi k}{q_{\phi }}=E^{\textrm{th}}=Td^{\textrm{th}}\geq k\frac{me|v|}{2}\frac{2\pi }{q_{\phi }}d^{\textrm{th}}\, ,\]
which results in 
\[
d^{\textrm{th}}\leq \frac{4}{me}\, .\]

\section{Summary and conclusions}

In this work we have extended for non-Abelian theories some of the ideas of
't Hooft and Mandelstam on quark confinement. Besides the fact that our theory
has an unbroken non-Abelian gauge group, another interesting feature is that
it has solitonic monopoles instead of Dirac monopoles which appear in the Abelian
theories. We have considered \( N=2 \) super Yang-Mills with arbitrary simple
gauge group, with one flavor and with an \( N=2 \) mass breaking term. We have
shown that, by continuously varying the mass breaking parameter \( m \), we
can pass from an unbroken phase to a phase with free monopoles and then to a
phase with \( Z_{k} \)-strings. This last phase occurs due to the fact that
the scalar \( \phi  \) acquires a non-vanishing expectation value. When \( k=2 \),
\( \phi  \) can be interpreted as a diquark condensate. We showed that the
magnetic fluxes of the monopoles is multiple of the fundamental \( Z_{k} \)-string
flux and therefore the monopoles can undergo confinement. We also obtained a
bound for the threshold length for a string to break in a new pair of monopole-antimonopole.
Following 't Hooft and Mandelstam ideas, one could expect that, in the dual
theory, with \( \phi  \) being a monopole condensate, it would happen a quark-antiquark
confinement. 

We have seen that some of our confining theories are obtaining by adding a deformation
to \( U(N) \) superconformal theories which breaks the gauge group further
to \( (SU(N)\otimes Z_{2})/Z_{2N} \) . It is expected that a confining theory
obtained by a deformation of superconformal gauge theory in 4 dimensions should
satisfy a gauge/string correspondence \cite{Witten} (which would be a kind
of deformation of the CFT/AdS correspondence \cite{Maldacena}). In the gauge/string
correspondences it is usually considered confining gauge theories with \( SU(N) \)
or \( U(N) \) completely broken to a discrete group. Therefore, it would be
interesting to know if those theories also satisfy some gauge/string correspondence.

\newpage

\vskip 0.2 in \noindent \textbf{}\textbf{\large Acknowledgments} \textbf{}

\noindent I would like to thank Patrick Brockill for reading the manuscript
and CNPq for financial support.

\end{document}